\documentclass{JAIS}

\received{xx January 2018}
\published{xx March 2018}

\def\be{\begin{equation}}
\def\ee{\end{equation}}
\def\bea{\begin{eqnarray}}
\def\eea{\end{eqnarray}}

\usepackage{caption}
\usepackage{lineno}
\usepackage{graphicx}
\articletype{Technical Report}
\usepackage[utf8]{inputenc} 
\usepackage[T1]{fontenc} 
\usepackage{gensymb}
\usepackage{mathpazo} 
\usepackage{latexsym}
\usepackage{booktabs}
\usepackage{etoolbox}
\usepackage{tikz}
\usepackage{xcolor}
\usepackage{multirow}
\usepackage{lscape}
 \usepackage{siunitx}
 \usepackage{threeparttable}
\usepackage{subfig}
\usepackage{caption}
\usepackage{adjustbox}
\begin{document}

\title{Small-area Portable Resistive Plate Chambers for Muography}

\author{A. Samalan\auno{1}, S. Basnet\auno{2},  E. Cortina Gil\auno{2}, P. Demin\auno{2}, R. M. I. D. Gamage\auno{2},  A. Giammanco\auno{2}, R. Karnam\auno{2,4}, V. Kumar\auno{2}, M. Moussawi\auno{2}, M. Tytgat\auno{1,3}, A. Youssef\auno{5}}
\address{$^1$Department of Physics and Astronomy, Ghent University, Ghent, Belgium}
\address{$^2$Centre for Cosmology, Particle Physics and Phenomenology (CP3), Universit\'e catholique de Louvain, Louvain-la-Neuve, Belgium}
\address{$^3$Physics Department, Vrije Universiteit Brussel, Brussels, Belgium}
\address{$^4$Centre for Medical and Radiation Physics (CMRP), National Institute of Science Education and Research (NISER) Bhubaneswar, India}
\address{$^5$Multi-Disciplinary Physics Laboratory, Optics and Fiber Optics Group, Faculty of Sciences, Lebanese University, Hadath, Lebanon
\\
\vspace{0.2cm}
Corresponding author: A. Samalan\\
Email address: amrutha.samalan@cern.ch, amrutha.samalan@ugent.be
}

\begin{abstract}
Muography is finding applications in various domains such as volcanology, archaeology, civil engineering, industry, mining, and nuclear waste surveys. To simplify transportation and installation in remote locations after laboratory testing, a fully portable and autonomous muon telescope based on Resistive Plate Chambers (RPCs) is being developed. Two glass-RPC prototypes have been created, sharing the same design goals but with different detector parameters, and comparative studies are ongoing. Drawing from prototype experience, a double-gap RPC with advanced features and improved spatial resolution is constructed. Resistive electrodes are produced manually, and a new data acquisition board is currently undergoing calibration. The results on prototype performance, readout board comparisons and the technical progress on the double-gap RPC are presented.
\end{abstract}

\maketitle

\begin{keyword}
Muon Radiography\sep Particle Detectors\sep Gaseous Detectors\sep Resistive Plate Chambers
\end{keyword}

\section{Introduction}

Muography is an imaging technique based on the absorption or scattering of atmospheric muons, produced when primary cosmic rays interact with the upper atmosphere of our planet. 
Muons are elementary particles with a mass roughly 200 times larger than the electrons. They are insensitive to strong nuclear interactions and, due to their relatively large mass, at the energies that are typical of atmospheric muons at sea level, they have a low rate of energy loss by ionization and other electromagnetic processes. 
This gives atmospheric muons a very high penetration power (larger than any other charged particle and only second to neutrinos), which
makes muography an appealing technique for the radiography of large and dense objects. 

Applications of muography are becoming ubiquitous~\cite{IAEA2022,MuographyBook}, ranging across archaeology~\cite{procureur2023precise}, volcanology~\cite{muraves2022}, geosciences~\cite{Lechmann:2021brn}, nuclear waste characterization~\cite{weekes2021material}, border controls~\cite{barnes2023cosmic}, nuclear warhead verification in the context of non-proliferation treaties~\cite{morris2014horizontal}, and industrial monitoring and maintenance~\cite{arbol2019non}. 

Most muography projects are based on large-area detectors, e.g., in order to surround volumes of the size of a cargo container or to collect large statistics to study huge targets such as volcanoes or pyramids; moving such large detectors is very complex, and in most cases, it is not necessary. 
However, some important use cases demand detector set-ups that are small and portable. 
This is often the case in archaeological or geophysical explorations, where the detectors may need to be located in narrow and confined environments such as tunnels or underground chambers~\cite{moussawi2021portable}. 
Use cases for portable detectors can also be found in cultural heritage preservation, e.g. statues hosted in crowded museum spaces and which are too heavy or too fragile to be transported to a laboratory~\cite{Moussawi:2023zkd}. 
A portable muon detector is also necessary for a new type of positioning system based on cosmic muons~\cite{tanaka2023muometric,Varga:2023ybt}, for use in places (e.g. underground) where more standard navigation techniques such as GPS are unavailable.

Several types of particle detectors are suitable for muography~\cite{Bonechi:2019ckl}; among them, gaseous detectors are often chosen when real-time information is necessary (which excludes nuclear emulsions) and spatial resolution is important. Resistive Plate Chambers (RPC), in particular, have been utilised by several teams because of their trade-off between performance and cost~\cite{MuographyBook-RPCchapter}. 
This article reports on the status of a project that aims at the development of portable, compact and autonomous muon detectors based on RPC. More details about this project and its previous development steps can be found in Refs.~\cite{Wuyckens2018,Basnet2020,Gamage_2022,Basnet:2022cds,gamage2022portable}.

This paper primarily focuses on conducting a comparative study involving three distinct glass-RPC (gRPC) prototypes developed with slightly different characteristics. In Sec.~\ref{detectordevelopment}, you will find comprehensive information pertaining to the geometry and layout of these three prototypes. This section also provides detailed information on the design and construction specifics of key hardware components, such as resistive electrodes, readout boards, and detector frames. In Section~\ref{electronics}, the overview of the detector electronics, encompassing elements like the Data Acquisition System (DAQ), HV modules, and more, is presented. For a comprehensive understanding of the data analysis and the significant outcomes derived from the comparison of gRPC prototypes, please refer to Sec.~\ref{analysisresults}.

\section{Detector Development}
\label{detectordevelopment}
\subsection{Chamber Layout}
\begin{figure}[h]
    \begin{minipage}{0.5\textwidth}
        \centering
        \rotatebox{90}{\includegraphics[width=0.6\linewidth]{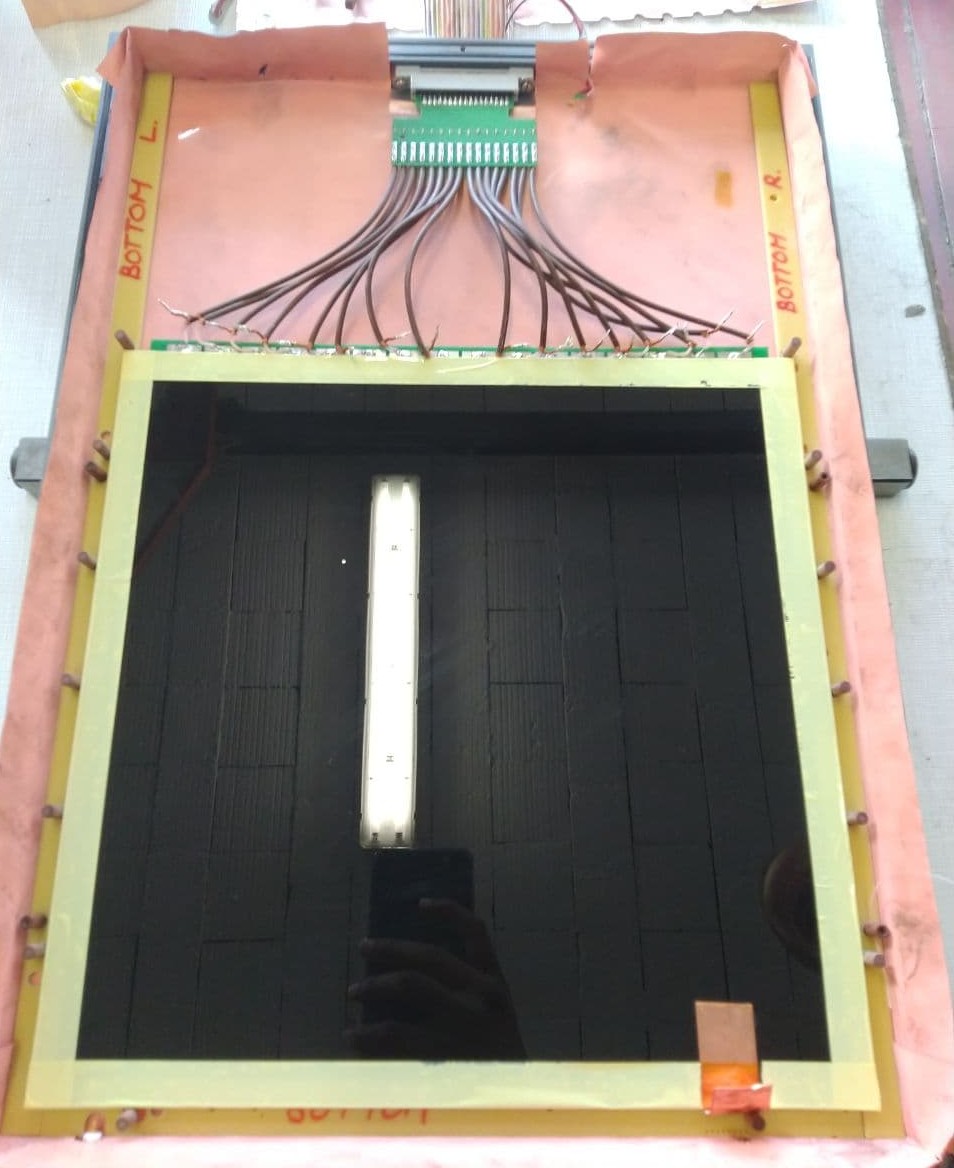}}
        \caption*{\centering{(a) gRPC-A1}}
    \end{minipage}
    \begin{minipage}{0.4\textwidth}
        \centering
        \includegraphics[width=1\linewidth]{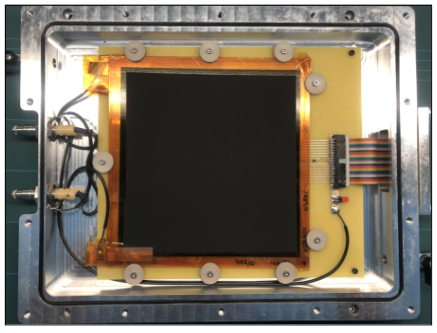}
        \caption*{\centering{(b) gRPC-B1}}
    \end{minipage}
    \\
    \begin{minipage}{\textwidth}
        \centering
        \includegraphics[width=0.35\linewidth]{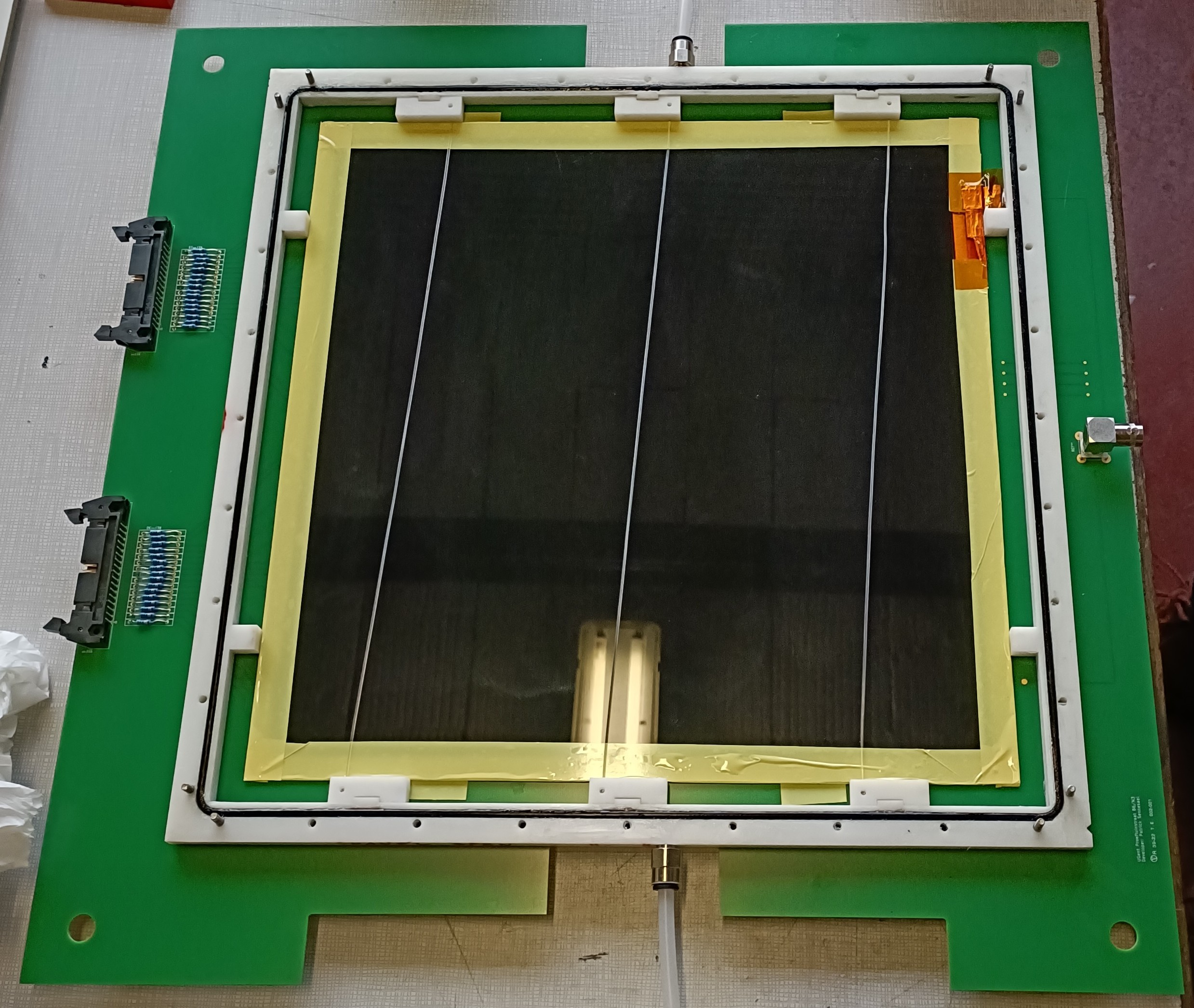}
        \caption*{\centering{(c) gRPC-C1}}
    \end{minipage}
    \caption{The inside view of the three gRPC prototypes developed.}
    \label{prototypepics}
\end{figure}

As part of the project, several gRPC prototype detectors were developed and tested. In order to validate the detector mechanics and design and the resulting performances, three prototypes with slightly different characteristics were constructed. 
Figure~\ref{prototypepics} shows the inside view of these prototypes.

\begin{figure}[h]
    \begin{minipage}{0.5\textwidth}
        \centering
        {\includegraphics[width=0.8\linewidth]{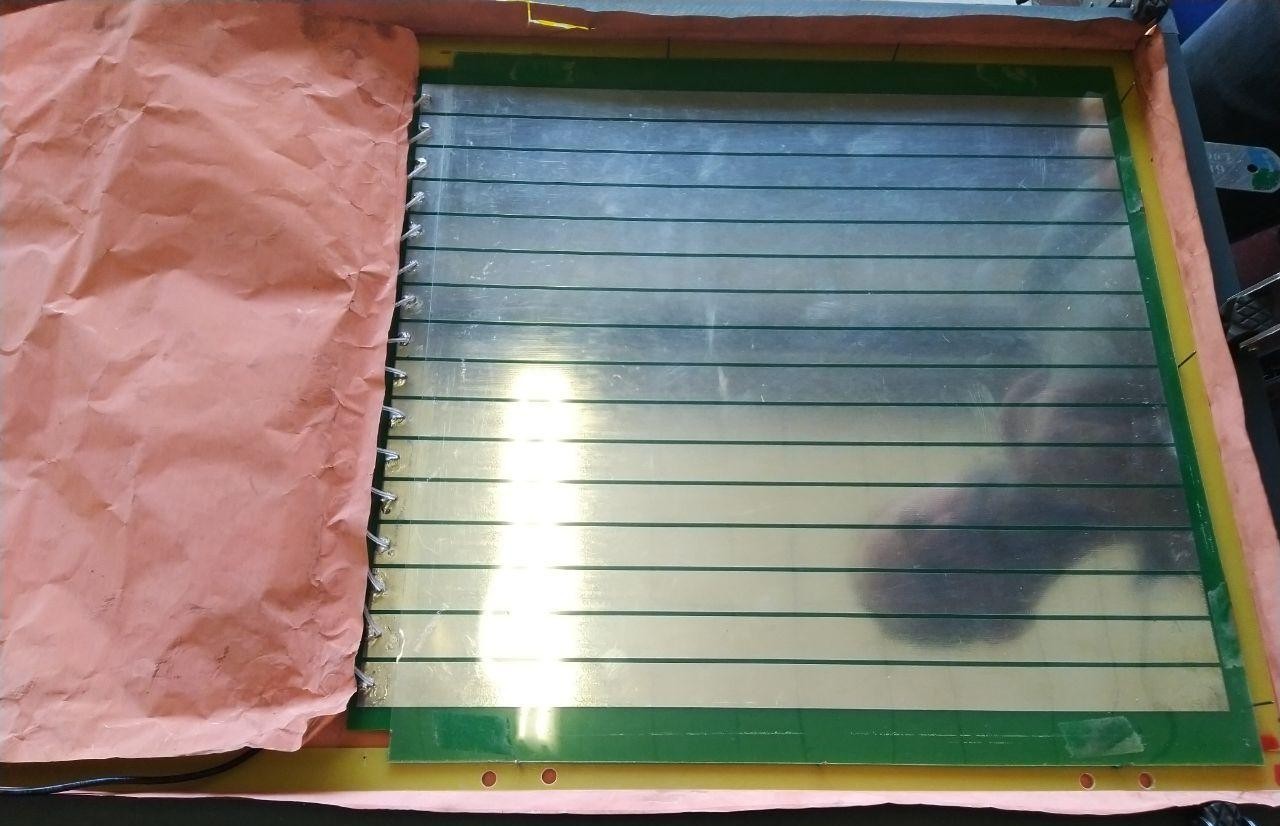}}
        \caption*{\centering{(a) PCB-A1}}
    \end{minipage}
    \begin{minipage}{0.5\textwidth}
        \centering
        \includegraphics[width=0.6\linewidth]{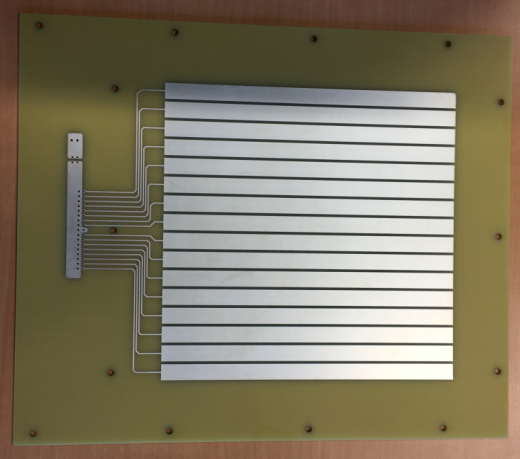}
        \caption*{\centering{(b) PCB-B1}}
    \end{minipage}
    \\
    \begin{minipage}{\textwidth}
        \centering
        \includegraphics[width=0.6\linewidth]{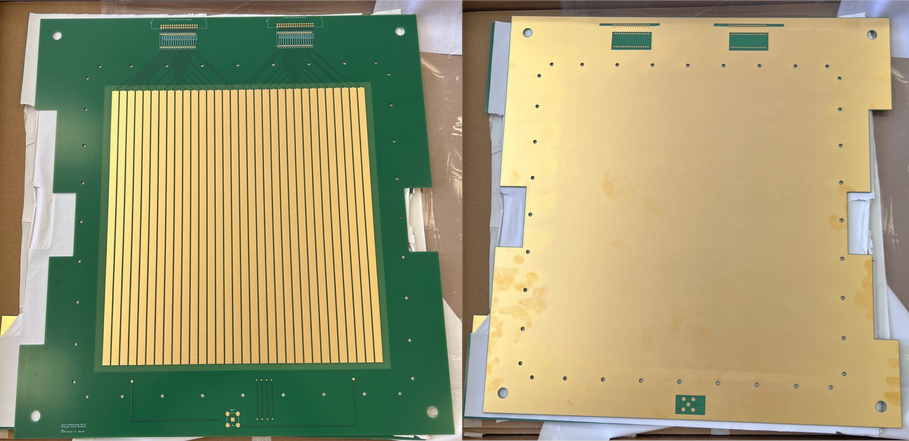}
        \caption*{\centering{(c) PCB-C1}}
    \end{minipage}
    \caption{The PCB boards used in the three gRPC prototypes developed. Images (a), (b), and (c) correspond to gRPC-A1, gRPC-B1, and gRPC-C1, respectively.}
    \label{pcbpics}
\end{figure}

The prototype gRPC-A1 is a basic single-gap RPC with an active area of 28$\times$28 $cm^2$. Here, three equally spaced nylon fish wires have been used to maintain a homogeneous gas gap of 1.0 mm between the 1.1 mm thick glass electrodes.
The resistive coating on the glass plates has been applied using an in-house spray-gun-based coating technique, achieving a uniform resistive layer in terms of both thickness and resistivity, which matches the quality level of industrially developed resistive plates. 
In this particular case, the plates have a mean surface resistivity of approximately 450 $k\ohm/ \Box$.
 The signal readout is facilitated by a Printed Circuit Board (PCB) containing 16 copper strips with a pitch width of 15 $mm$, separated by a distance of 1.6 $mm$ (Fig.~\ref{pcbpics}-(a)). 
 The glass plates, as well as the PCB, are placed inside a Faraday cage constructed with copper foils for noise reduction and the entire setup is encompassed within a lightweight gas-tight box based on a honeycomb structure.
 This box includes a high voltage (HV) connector, gas inlet, gas outlet, and a slot for connecting the flat cables extending outside the box to the readout electronics. Gas circulation is maintained within the system by a continuous low-flow rate of fresh gas mixture.
 Figure~\ref{prototypepics}-(a) provides an inside view of the prototype, showcasing the construction of the Faraday cage with copper foil, the resistive glass plate and the signal cables that extend from the PCB.

The gRPC-B1 is a single-gap RPC with a slightly modified detector layout compared to the initial prototype. It features a 16$\times$16 $cm^2$ active area while maintaining the same glass plate thickness and gas gap as the gRPC-A1. Unlike the gRPC-A1, which used fish wires, the gRPC-B1 employs nine circular round spacers made of polyether ether ketone (PEEK) placed along the edges of the glass plates. This design ensures that the entire resistive coated area remains active, eliminating any dead spaces. The resistive plates are of surface resistivity of approximately 4 $M\ohm$ and were produced using a distinct industrial coating technique called serigraphy, also known as silk screen printing, allowing for a direct comparison between the two coating methods. The PCB, in this case, also consists of sixteen copper strips, each with a reduced width of 1 $mm$ and separated by a 1 $mm$ gap (Fig.~\ref{pcbpics}-(b)). The resistive plates and PCB are hosted in an air-tight aluminum box, and the box incorporates slots for HV connections and flat cable connectors for both power supply and signal readout. The box is constructed with strategically placed O-rings of Nitrile rubber within the detector frame to ensure gas tightness. In vacuum conditions, the gas leakage rate was measured using helium gas, yielding an estimated value of 10$^{-9}$$mbar~l~s^{-1}$~\cite{Wuyckens2018}. In Figure~\ref{prototypepics}-(b), the internal arrangement of the detector within the aluminum enclosure is depicted. Notably, the image shows the circular spacers positioned along the edges of the glass plates, as well as the O-rings embedded in the frame.

The third prototype is an advanced version developed, benefiting from the lessons learned from the aforementioned ones. It is a double-gap detector offering a two-dimensional readout system. The detector's active area measures 28$\times$28 $cm^2$, with electrode and gas gap dimensions identical to those in the prior prototypes. The gas gaps, in this instance, are constructed using Nylon fish wires, with three placed equidistantly within each gap. Four glass plates, each with a surface resistivity of approximately 1.5 $M\ohm/ \Box$, construct the double-gap. The improved design features a PCB with 32 integrated 0.8 $cm$ wide strips, eliminating the need for cables by embedding copper traces on the PCB for signal guidance. Additionally, traces for high-voltage, ground connection, and the P-T-H sensor (pressure, temperature, and humidity sensor) are integrated into the PCB. This PCB also hosts 34 slots for resistors for 50 $\ohm$ termination to eliminate reflections. Importantly, the PCB itself acts as a cover for the detector, with both PCBs placed at the top and bottom of the detector frame, which also functions as an enclosure for component assembly. The detector frame is 3D printed with slots only for gas connectors. Therefore, once the box is closed, all connections from inside come out through the PCB traces. This design choice ensures complete gas-tightness, eliminating the need for slots in the frame, as seen in previous iterations. Both PCBs are positioned orthogonally at the top and bottom, providing two-dimensional information. Each PCB offers two slots for mounting the 17-pin connector, enabling a direct connection to the signal PCB traces. The backside of the PCB is coated with copper to establish a Faraday cage. Finally, copper foil seals all four sides of the detector, connecting the top and bottom layers to enclose the Faraday cage completely. In Figure~\ref{prototypepics}-(c), the layout of the double-gap chamber is depicted, featuring the 3D-printed frame, the resistive plates, the construction of the gas gap with fish wire, and various connectors on the PCB. The detailed picture of the PCB  board (front and back side view) is shown in Fig.~\ref{pcbpics}-(c).

The major parameters of all three prototypes are summarised in Tab.~\ref{summarytable}

\begin{table}[]
\resizebox{\textwidth}{!}{
\begin{tabular}{@{}c|ccc@{}}
{\textbf{Detector}} & {\color[HTML]{000000} \textbf{gRPC-A1}} & {\color[HTML]{000000} \textbf{gRPC-B1}} & {\textbf{gRPC-C1}} \\ \hline

\textbf{Active Area} & 28$\times$28 $cm^2$ & 16$\times$16  $cm^2$ & 28$\times$28 $cm^2$ \\

\textbf{Box type} & Honeycomb based & Aluminum casket & Closed with top and bottom PCBs \\

\textbf{Readout strips} & 16-1D & 16-1D & 32$\times$32 - 2D \\

\textbf{Strip pitch} & 15 $mm$ & 10 $mm$ & 8 $mm$ \\

\textbf{Gas gap} & \begin{tabular}[c]{@{}c@{}}1 $mm$\\ Single gap\end{tabular} & \begin{tabular}[c]{@{}c@{}}1 $mm$\\ Single gap\end{tabular} & \begin{tabular}[c]{@{}c@{}}1 $mm$\\ Double gap\end{tabular} \\

\textbf{Thickness of electrodes} & 1.1 $mm $& 1.1 $mm$ & 1.1 $mm$ \\

\textbf{Resistive coating} & Spray-gun ($\sim$650 $k\ohm/ \Box$) & Serigraphy ($\sim$4 $M\ohm/ \Box$) & \begin{tabular}[c]{@{}c@{}}Spray-gun ($\sim$1.5 $M\ohm/ \Box$)\end{tabular} \\

\textbf{Gas mixture} & \multicolumn{3}{c}{95.2\% Freon, 0.3\% SF6, 4.5\% isobutane} \\

\textbf{Portability} & No & Yes & Yes \\ \bottomrule
\end{tabular}
}
\caption{Summary of the components of the three detector prototypes developed for this study.}
\label{summarytable}
\end{table}

\subsection{Restive coating}
\begin{figure}[htbp]
\centering
\includegraphics[width=16pc]{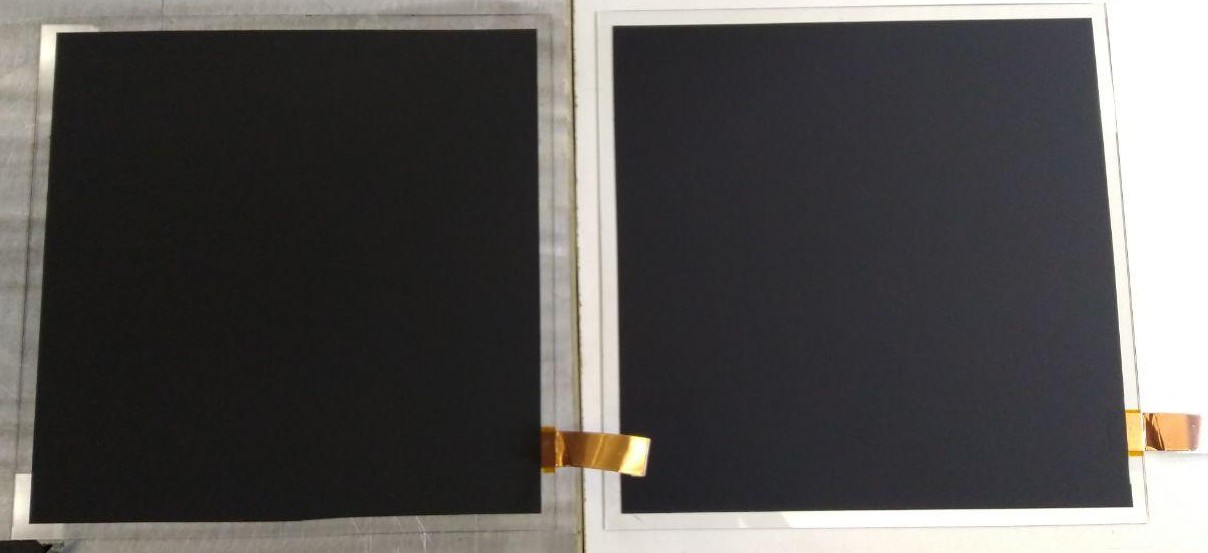}
\includegraphics[width=16pc]{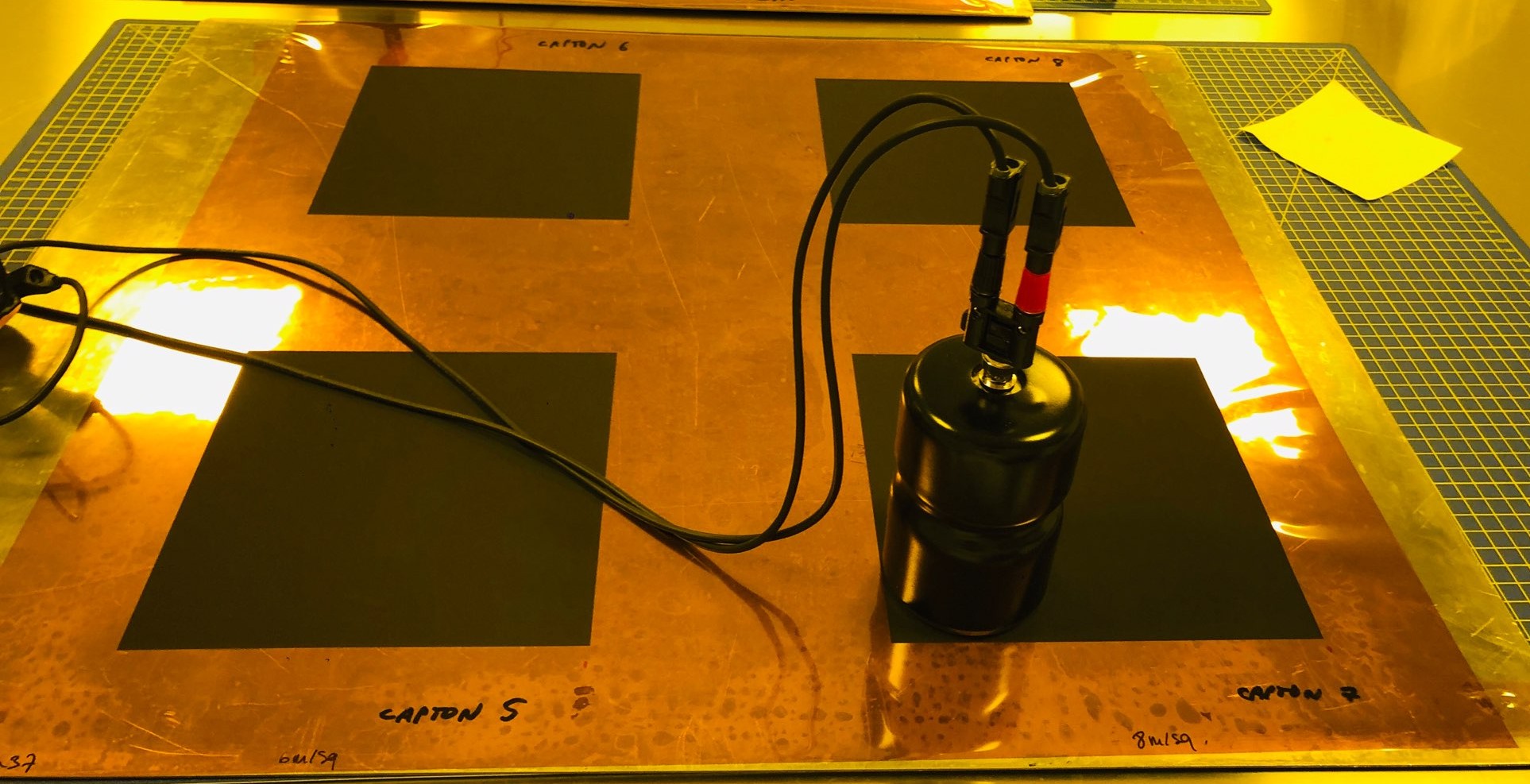}
\caption{Resistive plates developed for the prototype detectors using different coating techniques. The left side shows plates created using the spray-gun-based method, while the right side displays plates printed using the serigraphy technique. Both methods resulted in uniform coating thickness and surface resistivity with slight variations of up to approximately 15\%.}
\label{printedplates}
\end{figure}

\begin{figure}[h!]
\centering
\includegraphics[width=23pc]{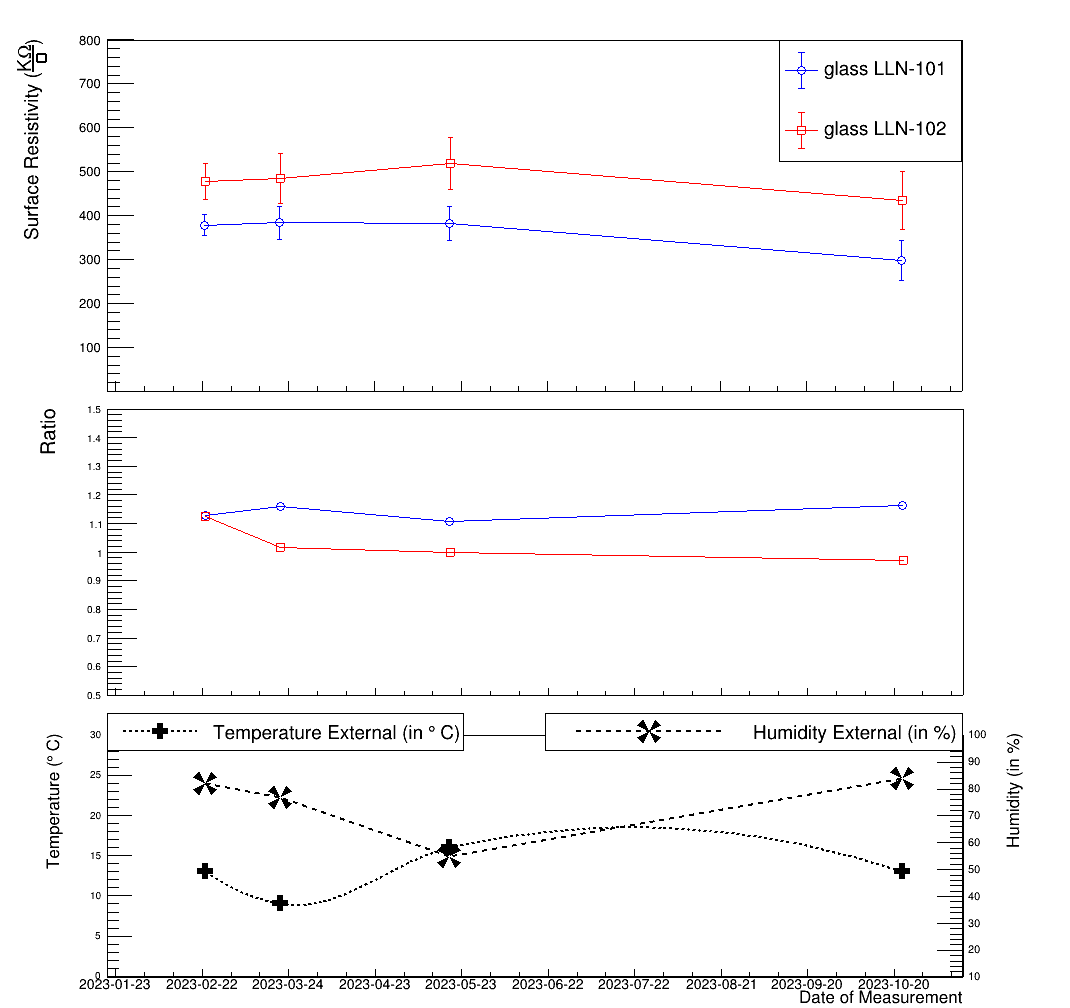}
\caption{Top: surface resistivity measurements over time for two plates manufactured using the spray-gun-based technique. Middle: The ratios of the surface resistivity measurement taken at two different locations (center vs. edge) of both plates as a function of the measurement date. Bottom: data on temperature and humidity recorded during the measurement period. Fluctuations in surface resistivity measurements can be attributed to variations in these environmental parameters.}
\label{surf_r_meas}
\end{figure}

As mentioned in the preceding section, two distinct coating techniques were investigated for the construction of the resistive plates. For gRPC-A1 and gRPC-C1, an in-house method was employed, involving the application of a blend of resistive and conductive paints onto the plates. This was achieved using a spray gun connected to a compressor, ensuring the requisite pressure (approximately $\sim 5$ bar). Subsequently, the freshly coated plates were subjected to baking in an oven at 110°C. The surface resistivity of the plates was then continuously monitored over several days to track variations. In the case of the second prototype, gRPC-B1, serigraphy was utilized. This technique entails the creation of a stencil to apply paint onto the glass surface. Figure.~\ref{printedplates} displays the resistive plates developed using the spray-gun-based method (on the left) and the serigraphy (on the right).
Both coating techniques resulted in uniform thickness and surface resistivity, exhibiting variations of approximately $\sim$15\%. Although some fluctuations in measurements were initially observed in the days following the coating in both cases, stability was subsequently achieved. Figure~\ref{surf_r_meas} (top) presents the analysis of surface resistivity measurements over time for two plates produced using the spray-gun-based technique. This graphical representation showcases the remarkable stability of surface resistivity, which has remained consistently within the same range even after 8 months since the initial coating. To assess the uniformity of the coating's surface resistivity, measurements were taken at various locations on the plates, and the resulting ratios (center vs. edge) are depicted as a function of the measurement date in Fig.~\ref{surf_r_meas} (middle). These ratios, ranging between 1.2 and 1.1 for both plates, indicate nearly identical surface resistivity across different regions. Moreover, as the resistivity measurements have exhibited minimal changes over time, the ratios also reflect a stable trend. For a comprehensive understanding of these measurements, Fig.~\ref{surf_r_meas} (bottom) includes data on temperature and humidity recorded during the measurement period. Fluctuations in surface resistivity measurements can be attributed to variations in these environmental parameters. Notably, in the surface resistivity plot, the slight increases observed in the case of the third data point coincide with a minor temperature rise and humidity decrease. Similarly, the slight decrease observed between the last two points aligns with an increase in humidity and a decrease in temperature over time.

In the case of some plates, due to prolonged exposure to the external environment, cracks in the coating were observed. To address this issue, a layer of urethane spray was applied to all plates post-development, which effectively prevented the crack formation. Even though both methods gave us similar results, it was decided to go with the spray-gun method for future work since this can be done in the laboratory itself manually, and the production cost is much cheaper.
\subsection{3D printed frames}
To ensure complete gas-tightness, the detector frame in the third iteration, the double-gap prototype, is produced as a single 3D printed piece. This 3D frame was developed as part of another in-house project and the idea was adapted for the construction of the double-gap prototype~\cite{samalanieee}. Unlike previous versions where significant gas leakage occurred at frame junctions and cabling slots, the improved PCB layout and the absence of any slots except for tiny ones for the gas connectors (diameter=2 mm) contribute to a fully sealed detector. The 3D-printed frame incorporates channels for accommodating O-rings, screw holes for connecting the top and bottom PCBs, and spacers to securely hold the glass plates in position. Additionally, the frame features an internal channel for uniform gas distribution when the chamber operates with gas flow.

\begin{figure}[htbp]
\centering
\includegraphics[width=16pc]{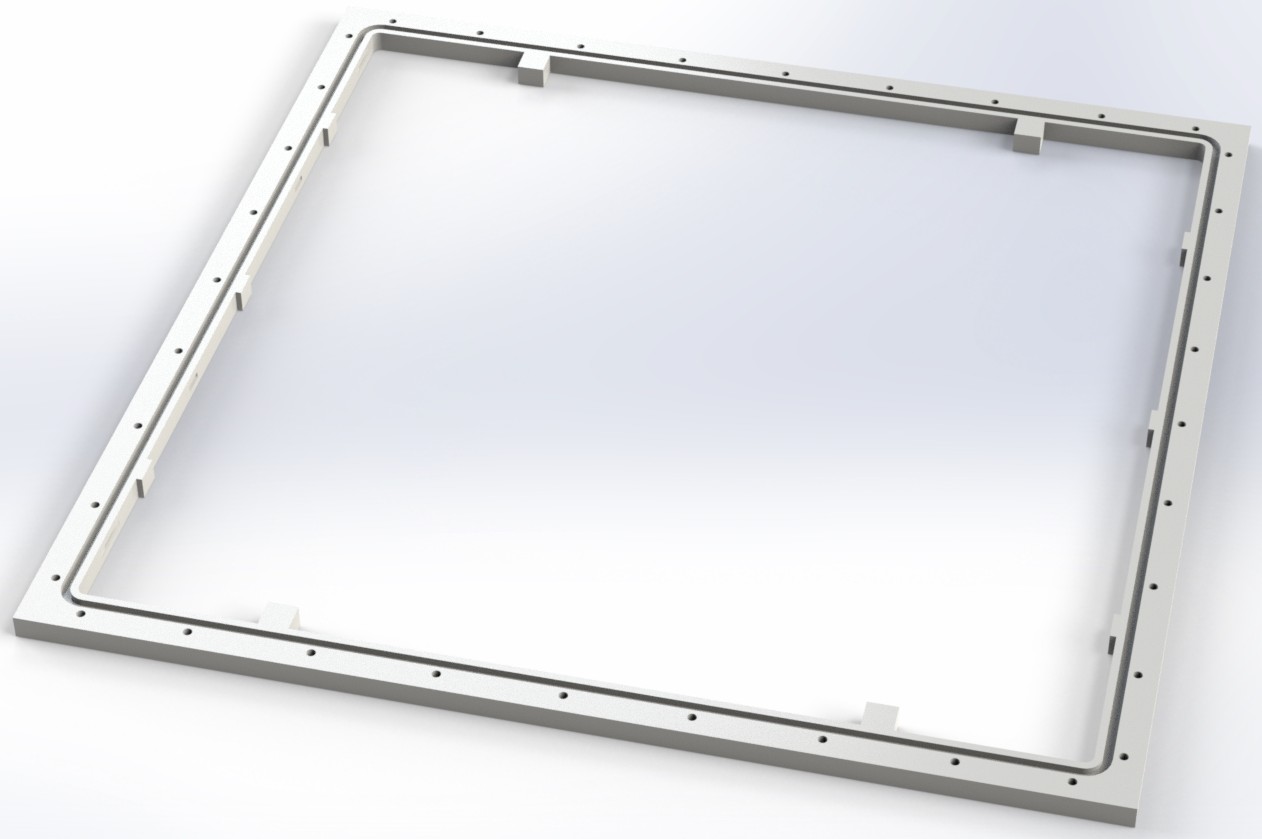}
\includegraphics[width=15pc]{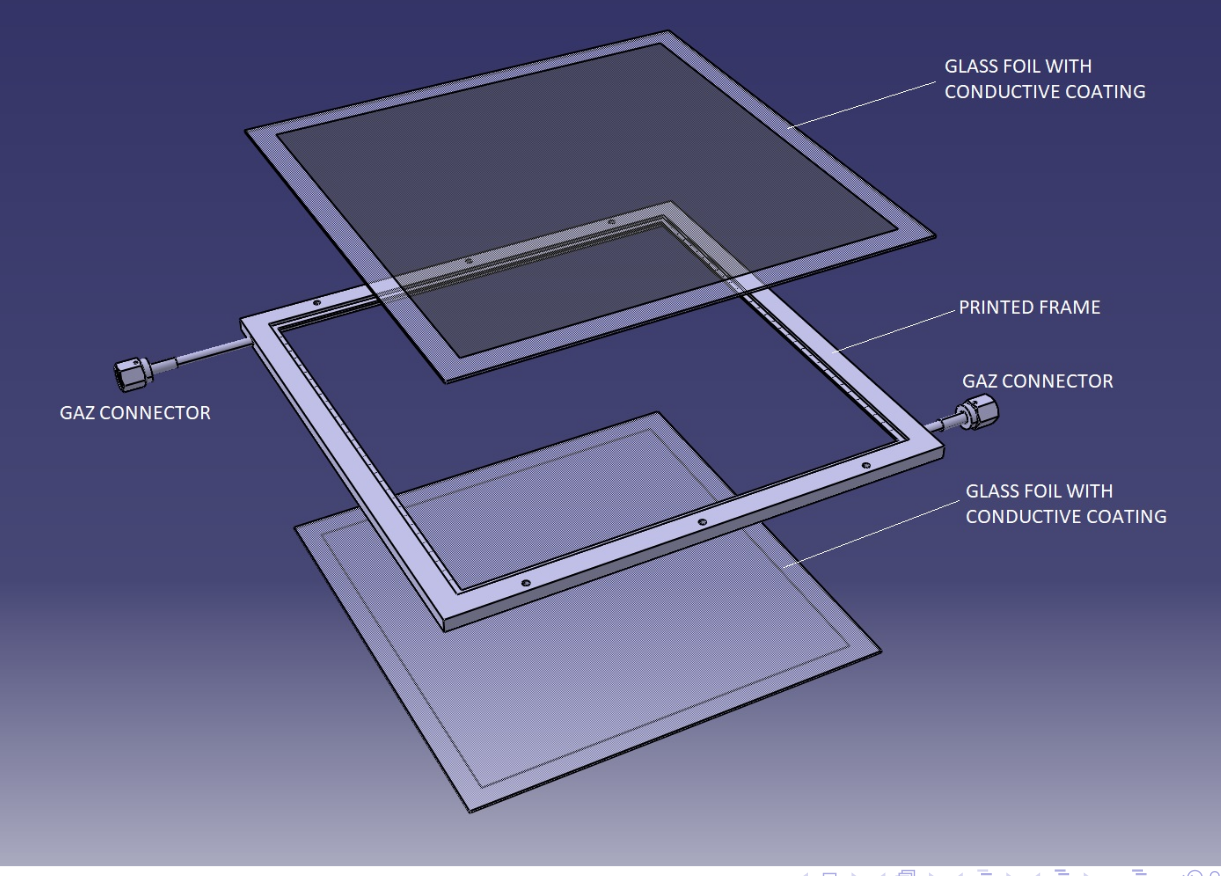}
\caption{CAD models of the 3D-printed frames, featuring the frame currently assembled in the gRPC-C1 (left) and the newly tested frame (right).}
\label{detectorframes}
\end{figure}

For detector configurations like gRPC-B1, where the entire detector box is filled with gas, an alternative frame design is employed, which is still in the testing phase. This frame is specifically designed to encase the glass plates within the box, and gas channels printed on the frame ensure uniform gas distribution while minimizing the required gas volume by directing flow only between the glass plates. The CAD models of the 3D-printed frame, currently in use in the gRPC-C1 (left), and the new frame being tested (right) are depicted in Figure~\ref{detectorframes}.
\section{Electronics}
\label{electronics}
A portable custom Data Acquisition System (DAQ) has been developed to facilitate data collection, ensuring complete portability. Utilizing two front-end boards (FEBs) borrowed from the Muon RPC system of the CMS experiment~\cite{FEB1, FEB2}, each FEB can read 32 analog channels, each containing an amplifier with a charge sensitivity of 2 $mV/fC$, a discriminator, a monostable circuit, and an LVDS driver. To guarantee autonomy, the LVDS outputs from all FEBs are linked to a System-on-Chip (SoC) module housed on a carrier board with wireless connectivity. Furthermore, the detectors are powered using a portable power supply module (Iseg DSP mini) capable of delivering HV up to 10 $kV$. Given gRPC-C1's advanced double-gap layout, additional measurements were carried out to assess the chamber's performance, including charge measurements. For this purpose, a setup based on CAEN QDC was employed. The setup featured a 32-channel CAEN QDC (V792) and two CAEN impedance adapters (A992), each accommodating 16 channels for 50 $\ohm$ impedance matching. In the case of all measurements, two scintillators of active area 16$\times$16 $cm^2$ are used to apply the trigger. The details and results obtained from all the conducted measurements are presented in the following section.
\section{Data Analysis and Results}
\label{analysisresults}
\begin{figure}[h!]
    \begin{minipage}{0.5\textwidth}
        \centering
        {\includegraphics[width=0.8\linewidth]{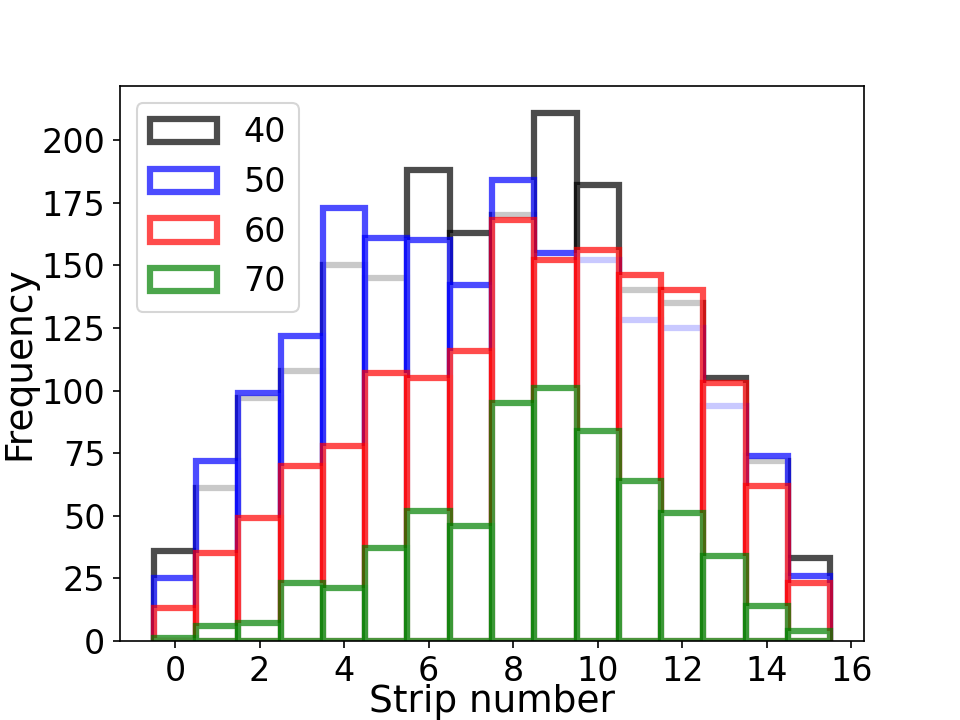}}
        \caption*{\centering{(a)}}
    \end{minipage}
    \begin{minipage}{0.5\textwidth}
        \centering
        \includegraphics[width=0.8\linewidth]{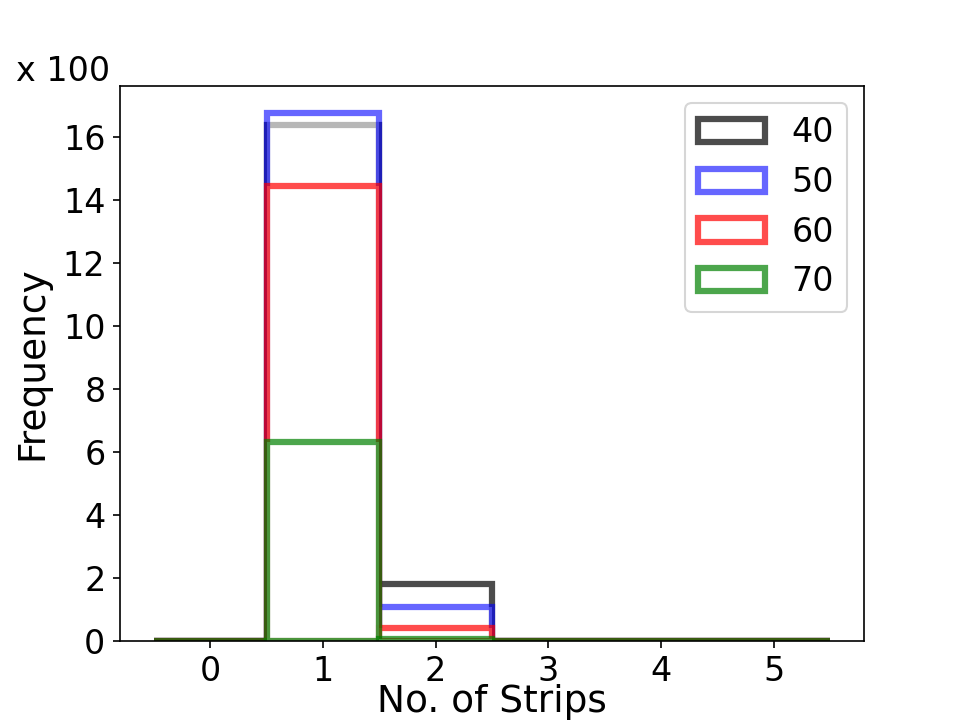}
        \caption*{\centering{(b)}}
    \end{minipage}
    \begin{minipage}{1\textwidth}
        \centering
        \includegraphics[width=0.5\linewidth]{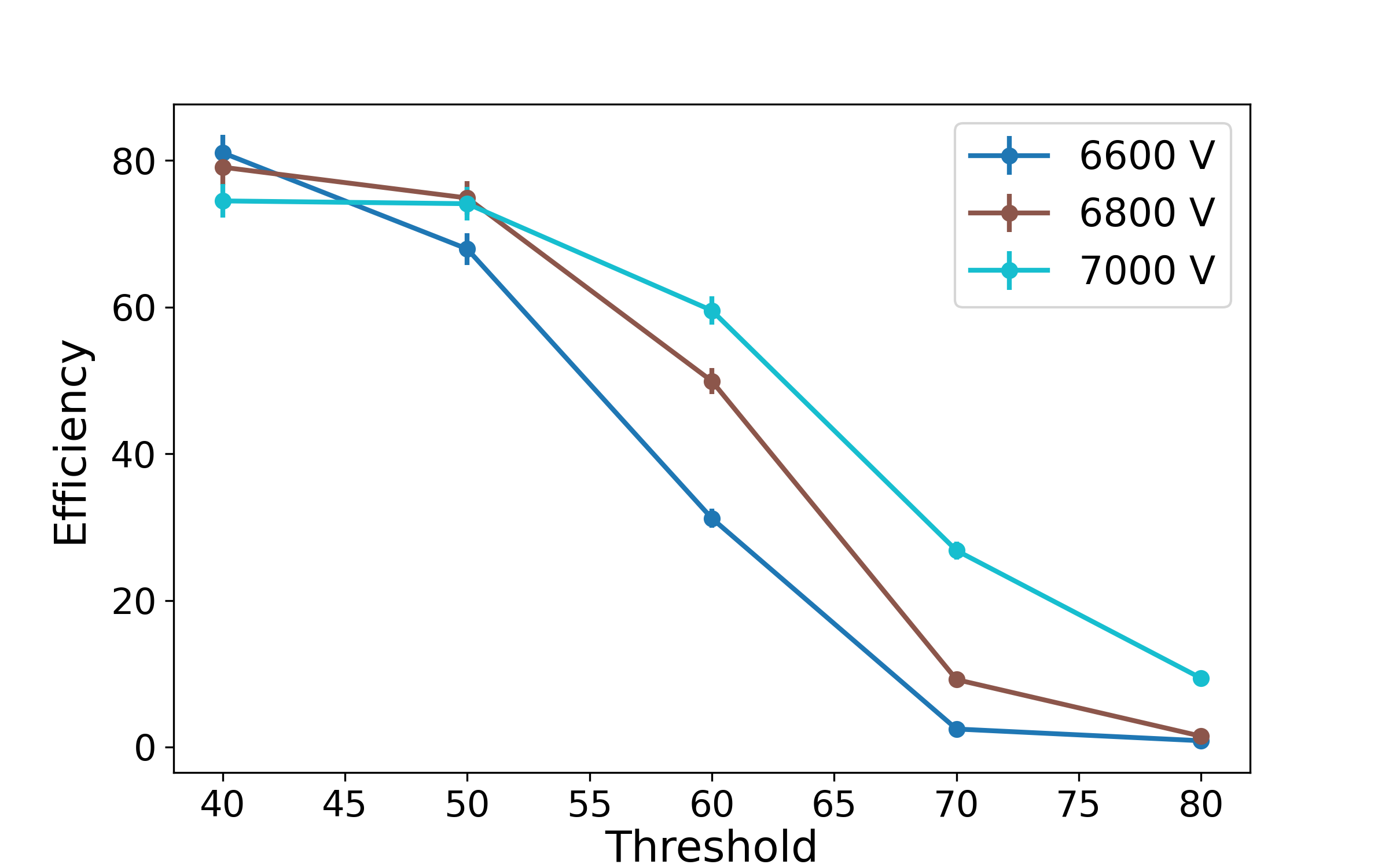}
        \caption*{\centering{(c)}}
    \end{minipage}
    \caption{Measurement results depicting (a) occupancy and (b) cluster size (number of strips fired per event). These measurements were conducted at HV 7 $kV$ with four different threshold values (in DAQ units). Additionally, (c) illustrates the variation in efficiency with respect to thresholds at three different working points (6.6, 6.8, and 7 $kV$).}
    \label{measurement1_results}
\end{figure}

\begin{figure}[h]
\centering
\includegraphics[width=24pc]{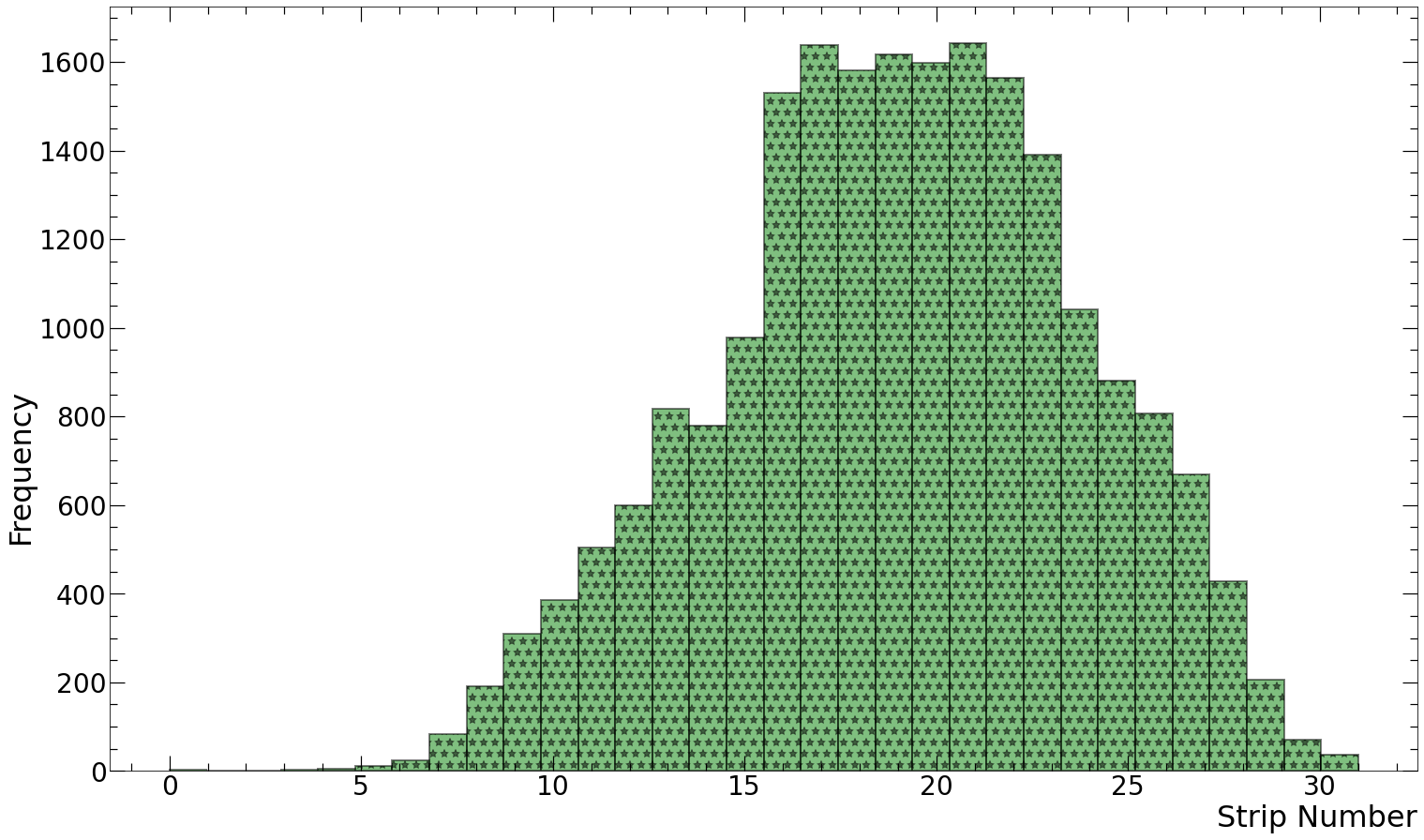}
\includegraphics[width=15pc]{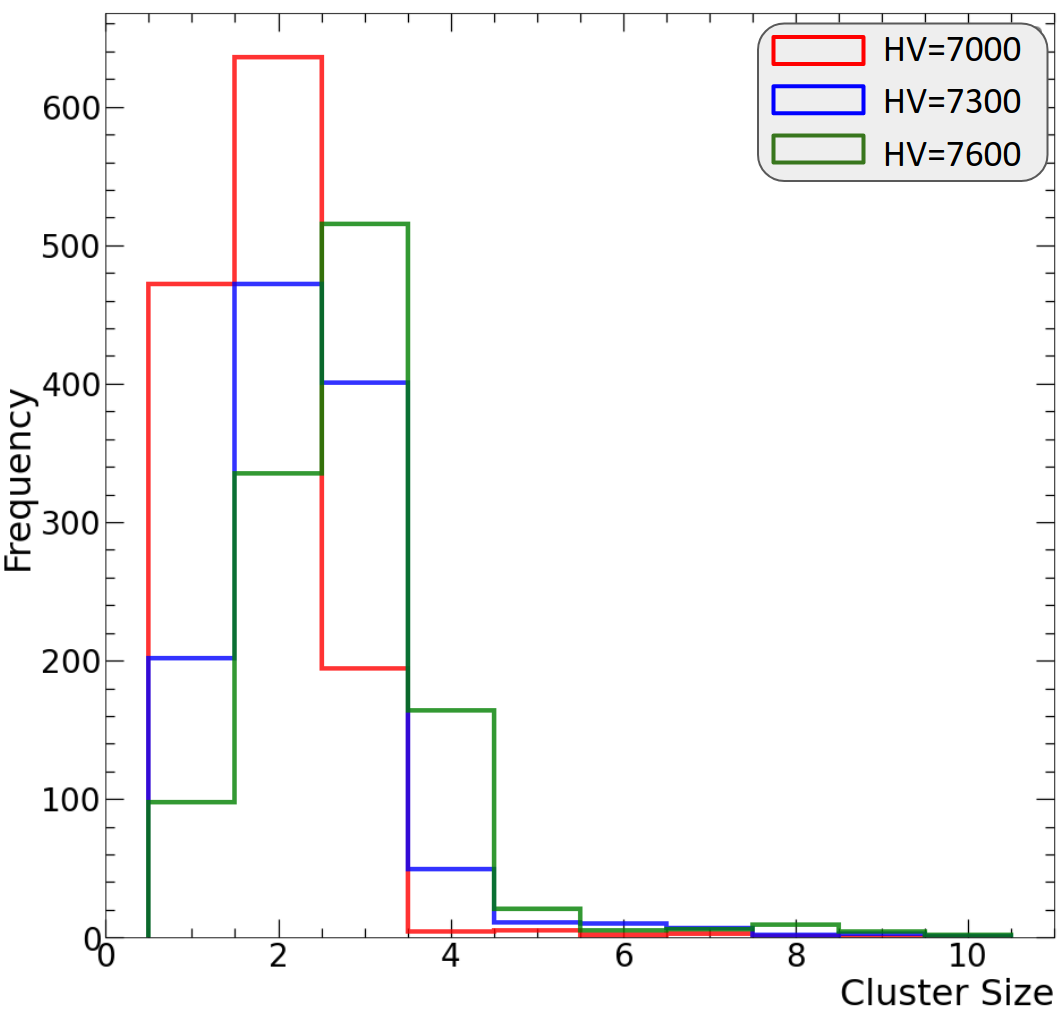}
\caption{First plot (left) shows occupancy distribution for the data collected with 10000 triggers. The lower values in initial strips are expected due to the difference between the detector active area and trigger scintillator size. The second plot demonstrates the cluster size distribution at three HV points (7, 7.3, and 7.6 $kV$).}
\label{meas2pics}
\end{figure}

A comprehensive set of measurements has been carried out to assess the performance of the developed RPC prototypes. These measurements encompassed an efficiency comparison study between the first two prototypes (gRPC-A1 and gRPC-B1) as well as an evaluation of key properties such as strip occupancy and cluster size. Detailed results of these comparative analyses can be found in~\cite{gamage2022portable}. Further investigations were undertaken to fine-tune the DAQ thresholds and compare cluster properties in both single-gap and double-gap configurations. In Figure~\ref{measurement1_results}, the results of these investigations conducted with gRPC-B1 are presented. The first plot (a) and the second plot (b) represent the occupancy and cluster size (number of strips fired per event) measurements, respectively. These measurements were conducted at four different threshold values (in DAQ units) and at HV 7 $kV$. The third plot shows the variation in efficiency with respect to the thresholds at three working points (6.6, 6.8, and 7 $kV$). 
As expected, both the occupancy level and cluster size decrease with increasing threshold values. Higher thresholds effectively reduce background hits, but the excessive setting can result in the elimination of potentially valuable events. To identify the optimal DAQ threshold value, efficiency concerning the threshold was studied. The results indicate that efficiency remains relatively stable up to 55 DAQ units for all three working points. Beyond this threshold, efficiency experiences a significant drop. Therefore, for subsequent studies, the threshold was set to values between 40 and 45 DAQ units to strike an appropriate balance.
The study of occupancy and cluster size in the case of the double-gap chamber, gRPC-C1, was also conducted, with the results displayed in Fig.~\ref{meas2pics}. In the first plot (left), the chamber's occupancy is depicted using data collected from 10,000 triggers. Notably, the occupancy is significantly low in the initial few strips due to the detector's larger active area compared to that of the trigger scintillators, resulting in certain strips remaining uncovered. The second plot illustrates the distribution of cluster sizes at three different working points (7, 7.3, and 7.6 $kV$). The plot clearly demonstrates an evident increase in cluster size as the high voltage (HV) is raised.
To perform the charge measurements, the CAEN QDC (refer to Sec.~\ref{electronics}) is utilized. Event selection is carried out based on a threshold applied to the QDC counts ($QDC_{thr}$). The determination of $QDC_{thr}$ for each detector channel is estimated from the pedestal run, following the equation:

\begin{equation}
QDC_{thr_i} = \mu_i + 3\sigma_i
\end{equation}

\begin{figure}[h]
    \begin{minipage}{0.5\textwidth}
        \centering
        {\includegraphics[width=0.7\linewidth]{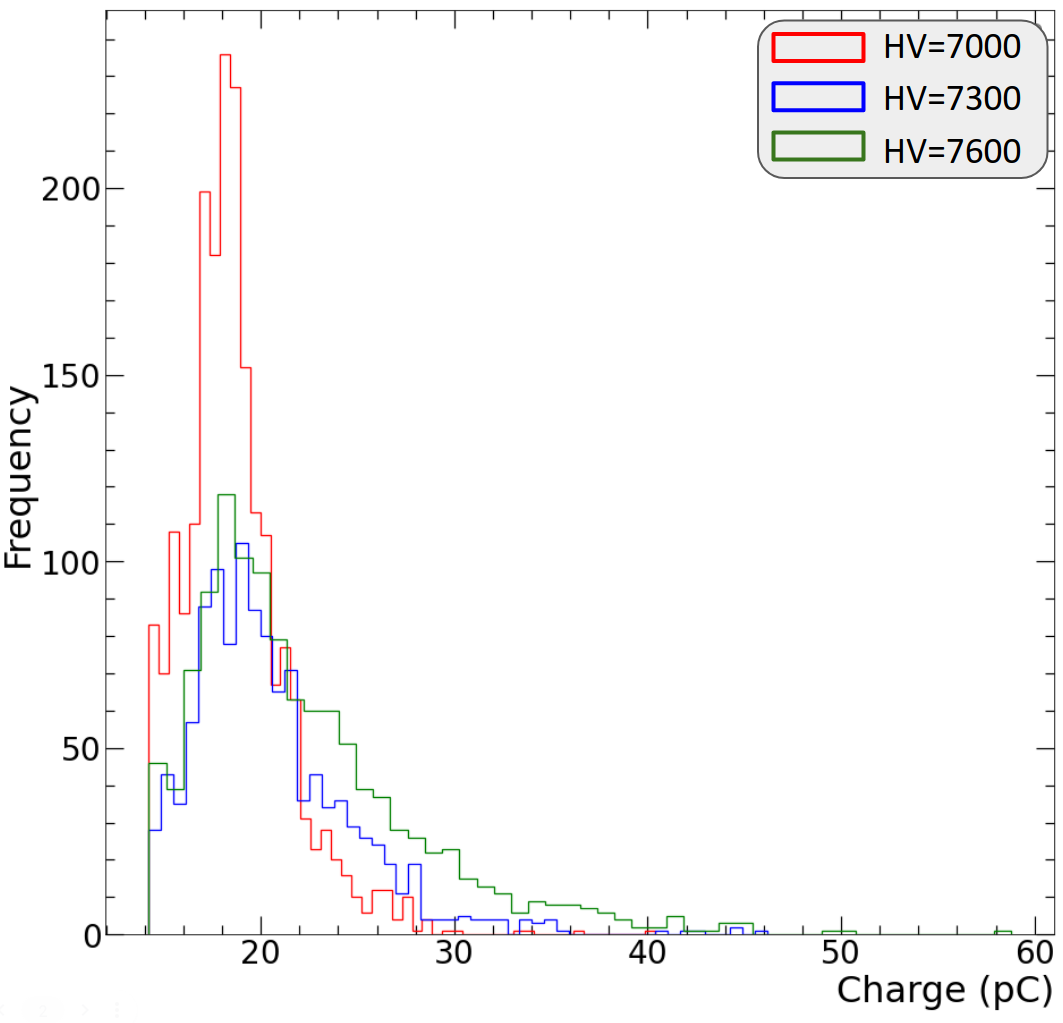}}
        \caption*{\centering{(a)}}
    \end{minipage}
    \begin{minipage}{0.5\textwidth}
        \centering
        \includegraphics[width=0.7\linewidth]{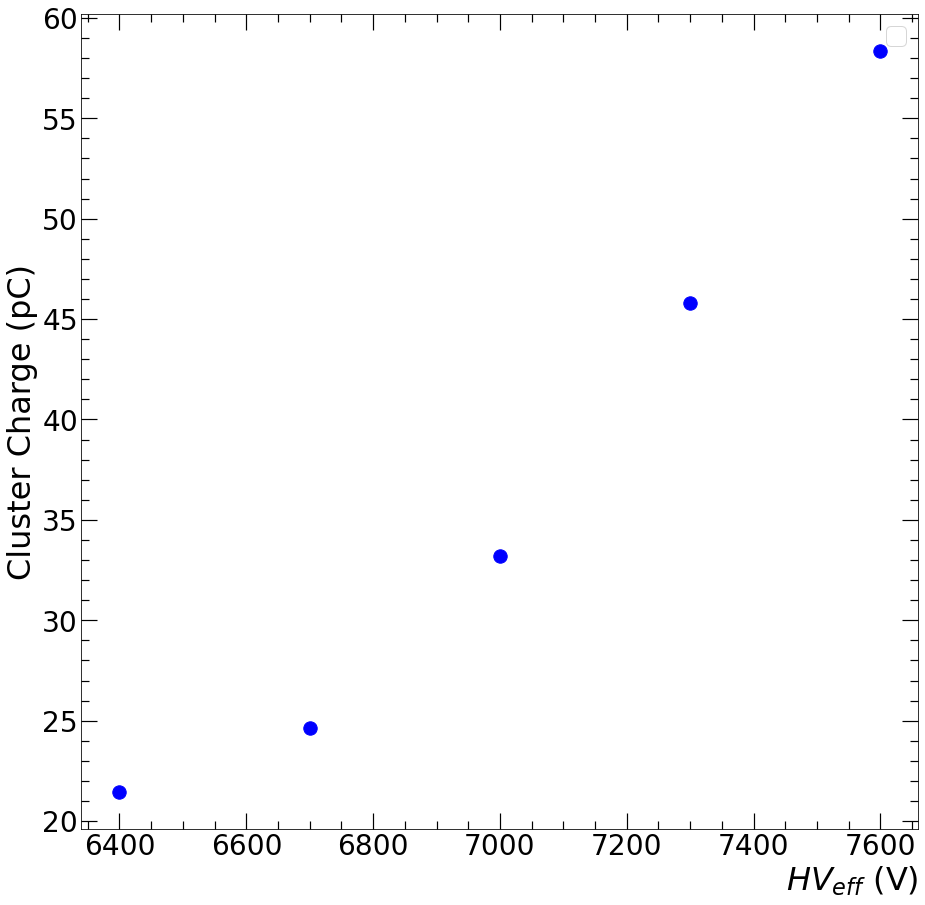}
        \caption*{\centering{(b)}}
    \end{minipage}
    \caption{ Plot (a) displays the charge measured in the detector at three different working points within the avalanche region. Plot (b) presents the average cluster charge as a function of HV.}
    \label{measurement2_results}
\end{figure}

In this equation, $QDC_{thr_i}$ is represented as the QDC threshold estimated for the $i^{th}$ channel of the detector. Here, $\mu_i$ denotes the mean QDC count associated with the $i^{th}$ channel in the pedestal run, and $\sigma_i$ is the standard deviation of the QDC count distribution for the $i^{th}$ channel. A strip is considered part of an event only if the QDC count surpasses the established threshold. The results of the charge measurement study are presented in Fig.~\ref{measurement2_results}.
In plot (a), the charge measured in the detector at three different working points within the avalanche region is displayed. It is evident that the charge increases with the high voltage (HV) applied. The charge per cluster is computed as the summation of the charge collected from the strips within the cluster. Plot (b) in Figure~\ref{measurement2_results} illustrates the average cluster charge as a function of HV.
\section{Conclusions}
This paper provides an overview of the development of portable muon detectors based on gRPCs for muography applications. Three different prototypes, gRPC-A1, gRPC-B1, and gRPC-C1, were developed with variations in design parameters and coating techniques. These detectors were designed to be compact, portable, and suitable for various applications, including archaeology, geophysics, and nuclear waste characterization. The paper discusses the design and construction details of these prototypes, including the resistive coating methods, 3D-printed frames, and electronics. Extensive measurements and comparative studies were conducted to evaluate the performance of these detectors, including efficiency, occupancy, cluster size, and charge measurements. The results demonstrate the suitability of these detectors for muography applications and highlight the importance of optimizing threshold values and high voltage settings to achieve the desired performance. Overall, this work contributes to the development of portable and efficient muon detectors for a wide range of applications, advancing the field of muography.
\section*{Acknowledgments}
This work was partially supported by the EU Horizon 2020 Research and Innovation Programme under the Marie Sklodowska-Curie Grant Agreement No. 822185, and by the Fonds de la Recherche Scientifique - FNRS under Grants No. T.0099.19 and J.0070.21. 
We gratefully acknowledge the work performed by Sophie Wuyckens between 2017 and 2019, which was crucial for the kick-off of this R\&D project. 
No conflict of interest exists. 

\bibliographystyle{unsrt}
\bibliography{refs}

\end{document}